\documentclass[a4paper,10pt]{article}

\usepackage{epsfig}

\begin{document}


\begin{center}
\noindent{\Large \bf Classification and discussion of macroscopic entropy production principles} \\

\vspace{15pt}

Stijn Bruers\footnote{email: stijn.bruers@fys.kuleuven.be, tel: 0032(0)16327503}\\
Instituut voor Theoretische Fysica, Celestijnenlaan 200D,\\
Katholieke Universiteit Leuven, B-3001 Leuven, Belgium\\
\end{center}

\begin{abstract}
In this article a classification of some proposed macroscopic entropy production (MEP) principles is given. With the help of simple electrical network models, at least six interesting and most used principles are distinguished: the least dissipation, the near-equilibrium (linear) minimum entropy production (MinEP), the near-equilibrium (linear) maximum entropy production (MaxEP), the far-from-equilibrium (non-linear) non-variational MaxEP, the far-from equilibrium variational MaxEP and the optimization MinEP. With this framework, the different assumptions, regions of validity, constraints and applications are explained, as well as their theoretical proofs,
counterexamples or experimental verifications. The examples will be kept as simple as possible, in order to focus more on the concepts instead
of the technicalities. By better defining the settings of the principles, this classification sheds some new light on some principles, and new ideas for future research are presented, especially for the more recent far-from-equilibrium principles. 
\end{abstract}

\footnotesize
\noindent {\bf pacs numbers:} 05.70.Ln, 65.40.Gr 
\normalsize

\section{Introduction}

The search for variational or extremization principles in physical systems is often quite fruitful. These can be applied to find the state of the system and its
stability \cite{Prigogine}, to describe (dynamical or static) fluctuations \cite{Maes}, to find dynamical laws, to find solutions
of these equations of motion, to find constraints on the direction of processes and evolutions,... In mechanics, one has the Lagrangian and Hamiltonian formalism with the principle of least action to find the
equations of motion, in equilibrium
thermodynamics there is the principle of maximum entropy or minimum free energy to find the equilibrium state, in near-equilibrium (linear) thermodynamics the entropy
production is minimized (MinEP) in order to find the stationary state,... \cite{Prigogine}

The situation far from thermodynamic equilibrium, and with nonlinear dynamics, is much more difficult. A general variational
principle is not known to exist, but can one at least define and describe regions where some principles do apply? The answer of this question not only involves thermodynamic constraints, but is also highly dependent on the kinetics (the balance or constitutional dynamical equations) of the system. As entropy and entropy production are such fundamental notions in irreversible thermodynamics, it is tempting to look for macroscopic entropy production principles (MEP). Besides the near-equilibrium MinEP, some of those proposed principles are the so called
the maximum entropy production (MaxEP) principles. 

The entropy production is used more and more to study physical systems, from simple electrical networks to complex chemical reaction systems, fluid systems or even ecological and climate systems. However, there is often confusion on MEP principles. This is not a coincidence, because e.g. the claim that both MinEP \emph{and} MaxEP
apply near equilibrium really sounds paradoxical. Furthermore, not always are the different principles clearly distinguished (as in the review \cite{Martyuchev}) and one sometimes uses different words with the same meaning or the same word with different meanings.  This is quite often the case for the MaxEP principles. We will see an attempt to translate the EP principles that are used in vastly different systems to analogous EP principles in electrical network examples which can be described as simple as possible. With this attempt, one can more clearly see whether there are different MaxEP principles.

The major message of this article is that by classifying and distinguishing the principles that are in use, stating their assumptions, applications, characteristics, constraints and experimental status, this article might help to disentangle some knots and will give a clearer language. Hence, it might also improve future research by stating more clearly the most challenging problems unsolved, keeping in mind the slogan that the first problem in every
research is to find a correct language and statement of the problem.

The following sections describe the different macroscopic entropy production
principles, which can be classified roughly in at least six principles: near-equilibrium least dissipation, near-equilibrium MinEP, near-equilibrium MaxEP, 
far-from-equilibrium non-variational MaxEP, far-from-equilibrium variational MaxEP and optimization MinEP. We will explain these principles with the help of simple electric network models, but these principles can be naturally applied to other models such as heat transfer, chemical reactions,... In the final section we will present some other EP principles applied in a more microscopic description, and also other less important, invalid or less clear principles are presented.

\section{General description}

Let us first start the description more generally. In this article far from equilibrium means far from global equilibrium, but still very close to local equilibrium, so that we can still speak of local intensive quantities such as temperatures, potentials,... and use the framework of local thermodynamics \cite{Prigogine} (see also \cite{Reguera} for more discussion about local and far from equilibrium issues). The basic thermodynamic variables such as temperatures, concentrations, potentials,... of the considered system are denoted with $M_{\alpha}$, and they have a dynamics derived from balance equations $\frac{dM_\alpha}{dt}=\sum_\beta J_{\alpha\beta}(M)$, where the right hand side contains the thermodynamic fluxes (also called currents or rates) $J_{\alpha\beta}=J_i$ ($i=1..N$ with $N$ the number of independent fluxes). The dynamics is called linear if $J_i(M)=\sum_\alpha j_{i\alpha}M_\alpha$.

Next, one can go to other variables, called thermodynamic forces, and these are functions of the basic variables: $X_i=X_i(M)$. Note that these forces are well defined for systems in local equilibrium. In the examples in this article, one can use these relations to write down the dynamics as
\begin{eqnarray}
\frac{d X_i(M)}{dt}=g_i(J(M)). \label{dynamics}
\end{eqnarray}
These equations do not yet form an autonomous system. In order to solve this system, one need additional so called phenomenological constitutive equations relating the forces with the fluxes: $J_i(M)=J_i(X(M))$, with the condition that all $X_i=0$ when all $J_j=0$. The latter condition means that we can write $J_i(M)=\sum_j L_{ij}(X,M) X_j(M)$. When all $X_i(M)=0$, the system is in thermodynamic equilibrium with $M=M^{eq}$.

The first important thermodynamic rate quantity is the entropy production rate density $\sigma$. It is a function of the thermodynamic forces and the thermodynamic
fluxes, and more specifically the bilinear form of forces and fluxes:
\begin{eqnarray}
\sigma = \dot{S}=\sigma(X,J)=\sum_i X_iJ_i, \label{defEP}
\end{eqnarray}
The total EP is the volume integral of this density. This is a nice quantity, because it is the product of kinetic quantities, the fluxes giving the time dependence of processes, and thermodynamic quantities, the forces giving the strength and direction of processes. If there is a microscopic entropy defined, one can derive the above EP expression \cite{Prigogine}. 

Another important quantity is the dissipation rate density or dissipation function, which is only a function of the fluxes:
\begin{eqnarray}
D = D(J). 
\end{eqnarray}
Specific examples will be given below. Note that some authors have different definitions for the dissipation function, such as $T\sigma$ with $T$ the local temperature field (e.g. \cite{Paltridge3}), or $T_0\sigma$ with $T_0$ the temperature of the environment or the temperature the system would acquire if it is in thermodynamic equilibrium with the environment (e.g. \cite{Jorgensen}). These correspond with the local heat production rate and the exergy destruction rate (rate of lost work) respectively \cite{Demirel}. The different usages of the word 'dissipation function' sometimes causes confusion in the literature, but we in this paper we will not go deeply into this topic. Here we will stick to our convention, the same as in e.g. \cite{Onsager}.

there exists an important dissipation function due to Onsager, but in order to define this, we need some discussion about closeness-to-equilibrium. We will define close-to-equilibrium as the range of forces and fluxes where the phenomenological constitutive equation is a linear response law to a sufficient degree of approximation:
\begin{eqnarray}
J_i\approx\sum_j L_{ij}X_j, \label{J=LX} 
\end{eqnarray}
with $L_{ij}=\lim_{X \to 0, M=M^{eq}} \frac{\partial J_i}{\partial X_j}$ the constant (independant on $M$ and $X$) linear response matrix which can be shown to be symmetric and invertible \cite{Onsager}. Note however that a priori there can be systems with very large values of the forces, but still these systems can be in local equilibrium and even in the linear response regime if the balance equations that specify the fluxes are of the required form. We will call systems with high values of forces but still linear response properties also near-equilibrium. An important remark (causing sometimes confusion as well) is that we need to keep a distinction between linear dynamics and linear response. We will come back to this point later, when studying the far-from-equilibrium non-variational MaxEP principle.

With this, the near-equilibrium EP density equals the famous Onsager dissipation function which is a quadratic function of the currents:
\begin{eqnarray}
D_{Ons} \equiv \sum_{ij}L_{ij}^{-1}J_iJ_j. \label{defOnsdiss}
\end{eqnarray} 
In the non-linear-response regime the EP and the Onsager dissipation function diverge from each other. The importance of a dissipation function is explained in the near-equilibrium least dissipation principle in the next section. 

In this article we will use electric circuits as explanatory examples. They always have external voltages denoted with $E_{a}$ and a number of resistance $R_i$, placed in series or parallel. $J_{a}$ are the currents through the external voltages. $U_i/T_i$ is the force\footnote{These are the definitions of forces and fluxes we will use in this article. However, some authors have different ones, calling $U_i$ the thermodynamic forces. This is only matter of convention, because only the product of the forces and fluxes, the EP rates, are relevant (writing $XJ$ or e.g. $(TX)(J/T)$ is equivalent).}  over the $i$-th resistance at temperature $T_i$ with voltage $U_i$, and $J_i$ are the thermodynamic fluxes (electrical currents) trough the corresponding resistance. The linear response matrix is given in terms of the resistances by $L_{ij}=\delta_{ij}T_i/R_i$, so the system is in the linear response regime when $R$ is independant on time or other variables. 

The stationary states (if they exist) are given by the solution of the dynamical equations, given by the set of mass, energy, entropy, electrical charge,... balance equations, together with some phenomenological constitutive equations and with some constraints. We will only use the charge degrees of freedom and not the electromagnetic field, and hence a simple electrical network with only resistors will have only a stationary state, and no transient states. We will introduce a non-trivial transient dynamics by placing capacitances with constant capacities $C_i$ in parallel with the resistances. The dynamics (\ref{dynamics}) can now be written as $\frac{d U_i}{dt}=f_i(C, J)$. Explicit expressions will be given below.

\section{Near-equilibrium least dissipation}

This principle was introduced by Onsager \cite{Onsager, Onsager2} as a generalization of work by Lord Rayleigh
\cite{Rayleigh} on fluid dynamics. It is often called the "principle of least dissipation", but this might sound confusing because it is not only the dissipation which is minimized. It states that
$\sigma(X,J)-D(J)/2$ is a maximum at solutions obeying the phenomenological constitutive laws. So the EP should be
as large as possible with at the same time the dissipation
as small as possible. It was the intention of Onsager and Rayleigh to derive phenomenological dynamical equations (like Navier-Stokes) with such an extremisation formalism, by looking for correct dissipation function. They have found that for near-equilibrium systems, with linear response, the correct dissipation function is a quadratic one, given by the Onsager dissipation function (\ref{defOnsdiss}). The search for general far-from-equilibrium dissipation functions is still open. 

The solution of this principle gives the phenomenological law, which is valid in the transient as well as in the stationary states. But there is more involved, which we will now briefly discuss (for more elaborate discussion: \cite{BruersMaesNetocny}). The principle is like the Lagrangian or Hamiltonian formalism, where a quantity is extremized in order to find the mean, macroscopic equations of motion. These equations of motion are only deterministic approximations in a thermodynamic/hydrodynamic limit. The usefulness of this least dissipation principle lies primarily in the deviations from this mean behavior, resulting in e.g. the fluctuation-dissipation theorem. If we write $\mathbf{P}(\omega(J(t)))\propto \exp \int dt \mathcal L(J(t))$ as the probability that the system will evolve along microscopic trajectory $\omega(J)$, with $\mathcal L(J)$ the Lagrangian, 
we get the stochastic phenomenological or dynamical equations. Let us apply this principle to an  electric network. In this principle no knowledge about the dynamics, the capacitances, the network structure, the external voltages and the external currents is necessary, so they can be arbitrary. Only the values of the resistances and the temperatures should be known.
Maximizing
\begin{eqnarray}
\mathcal L_U(J)&=&\sigma -\frac{D_{Ons}}{2}= \frac{U_1J_1}{T_1}+\frac{U_2J_2}{T_2}-\frac{1}{2}\left(\frac{R_1J_1^2}{T_1} + \frac{R_2J_2^2}{T_2}\right),
\end{eqnarray}
with respect to the flux variables $J_i$ at constant forces (hence the index $U$ in the notation of the Lagrangian) gives Ohm's law $J_{i,max}=U_i/R_i$. This law gives the constitutive equations (\ref{J=LX}) and is additional to the other equations of motion coming from balance equations (see (\ref{dynJ}) and (\ref{Energybalance}) in the next section). But due to fluctuations, we have the more general form $R_iJ_i=U_i+U_i^f$, with the fluctuating potential obeying the Nyquist formula $U_i^f=\sqrt{2R_iT_i}\xi_i$ with $\xi$ a standard white noise. 

\textbf{To summarize}: The near-equilibrium least dissipation uses the following assumptions: 
\begin{itemize}
\item the EP is given in terms of sums of products of forces and fluxes,
\item the dissipation is given in terms of sums of squared fluxes,
\item no equal temperatures are required,
\item no further constraints are required,
\item $\sigma -D/2$ is maximized keeping the forces constant, i.e. it is maximal in the system with the correct constitutive equation,
\end{itemize}
and it is used to find the phenomenological constitutive (linear Ohm's) laws (\ref{J=LX}), i.e. it puts restrictions on the equations of motion coming from balance equations. The position of the stationary state is not obtained. This extrimization principle is not true at a more microscopic, stochastic level; it only gives the mean behavior without fluctuations. Microscopically (i.e. not in the thermodynamic limit), there are deviations from the minimum.

The phenomenological laws thus obtained can become a starting point for the other near-equilibrium MinEP and MaxEP principles.

\section{Near-equilibrium MinEP}

The MinEP principle is perhaps the best known and has its origins in the work by Onsager \cite{Onsager} and Prigogine \cite{Prigogine}, and is
reviewed in e.g. \cite{Jaynes}. MinEP states that there's a unique
near-equilibrium\footnote{Equilibrium is always used to denote thermodynamic equilibrium, and should not be confused with stationarity.} stationary state in the presence of a number of fixed
external constraints (fixed combinations of thermodynamical driving forces or fluxes), and this state is given by the minimization of the entropy production under variation of
the non-fixed forces or fluxes. The non-fixed forces settle themselves in such a way that the conjugate fluxes vanish. A stronger version states
that near thermodynamic equilibrium the time derivative of the entropy production rate (EP) is always
negative. 

\begin{figure}[!ht]
\centering
\includegraphics[scale=0.7]{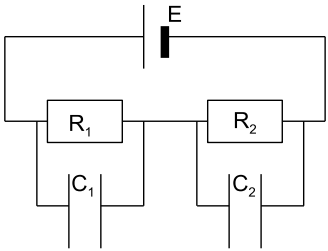}
\caption{The electrical network used in the explanation of the MinEP principle.}
\label{MinEPnetworkfig}
\end{figure}

Let us now consider a specific simple example (Fig. \ref{MinEPnetworkfig}): a circuit with two resistances $R_1$ and $R_2$ (plus parallel capacitors) placed in series and a fixed external voltage
(applied driving force) $E$. The electrical currents (thermodynamic fluxes) through the resistances are time dependent and written as $J_1(t)$ and $J_2(t)$, the temperatures of the resistances are $T_1$ and $T_2$ (and for simplicity are kept constant, i.e. we do not take backreaction due to heating into account), the thermodynamic forces over the resistances are $U_1(t)/T_1$ and $U_2(t)/T_2$ with $U_1$ and $U_2$ the voltages. Next we have to give the definitions of the basic quantities, as well as the dynamics.

The heat production, the Onsager dissipation function and the entropy
production are in the network example defined by:
\begin{eqnarray}
\frac{dQ}{dt} &\equiv& U_1J_1+U_2J_2, \\
D_{Ons} &\equiv& \frac{R_1J_1^2}{T_1} + \frac{R_2J_2^2}{T_2}, \\
\sigma &\equiv& \frac{U_1J_1}{T_1}+\frac{U_2J_2}{T_2}.
\end{eqnarray}
As mentioned before, near-equilibrium in the network means that the thermodynamic driving force, which is $E/T_0$ (with $T_0$ the temperature of the external environment) and also the other forces $U_i/T_i$ are small enough such that linear response is valid, resulting in Ohm's law for voltages and electrical currents:
\begin{eqnarray}
U_i = R_i J_i \label{Ohm}
\end{eqnarray}
with constant (not depending on time, temperature or other variables) resistances. So we take the currents obeying the near-equilibrium least dissipation principle. Furthermore, in near equilibrium we have to take the special and non-trivial restriction that the temperatures at the two resistances are nearly the same and close to $T_0$. If the temperatures were highly different, one would have a second external driving force at order $\Delta T$ which is not to be neglected. The final solution for the stationary states would differ in an order $E\Delta T$ which is one order higher than $E$. This was mentioned in \cite{Jaynes}. So with te nearly equal temperature approximation we get
$\sigma=\frac{dQ}{T_0dt}$, and so Joule's law follows: $dQ/dt=T_0D$. 

The charge balance equation (\ref{dynamics}) is:
\begin{eqnarray}
\frac{dU_1}{dt}&=&-\frac{dU_2}{dt}=a(J_2-J_1),\label{dynJ}
\end{eqnarray}
with $a=1/(C_1+C_2)>0$ the inverse sum of the two capacitances. The constitutive equations are again given by 
\begin{eqnarray}
U_1=R_1 J_1, \quad U_2 =R_2 J_2
\end{eqnarray}
(so from now on $\sigma=D_{Ons}$) and the constraint is
\begin{eqnarray}
E=U_1+U_2=\textrm{constant}. \label{consteq}
\end{eqnarray}
Note that this constraint is valid both in the transient and in the stationary states.

\begin{figure}[!ht]
\centering
\includegraphics[scale=0.7]{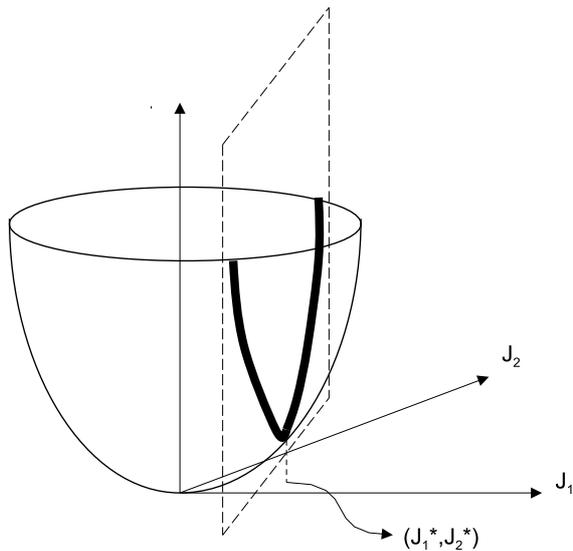}
\caption{Schematic plot of the MinEP principle. The thick parabola is the intersection of the quadratic EP potential with the vertical plane (dashed lines) as constraint. The minimum of this parabola gives the stationary state currents. The other points on this parabola are transient states.}
\label{classMinEPfig}
\end{figure}

The usefulness of MinEP is that there is another way to find the location of the unique stationary state (denoted with asterisks) without the use of (all of) the dynamical equations. The constraints and constitutive equations should be known, together with the correct expression of the EP. Finding the stationary state is now done with the help of Lagrange
multipliers $\lambda$. The function 
\begin{eqnarray}
\mathcal L\equiv \sigma + \lambda (E-U_1-U_2)
\end{eqnarray}
is extremized with respect to $J_1$ and $J_2$. The derivative with
respect to $J_1$ gives $\lambda=2J_1/T_0$, and with respect to $J_2$: $\lambda=2J_2/T_0$. So the stationary state is given
by $J_1^*=J_2^*$, i.e. Kirchoff's current law at the node between the resistances is obtained which states that charge is conserved. $L$ is now $L=2J_1E-(R_1+R_2)J_1^2$. Taking
again its derivative gives $J_1^*=E/(R_1+R_2)$. In the stationary state $\sigma$ is a minimum under the constraint. It can be visualized as the
parabolic intersection of a
quadratic potential paraboloid (the EP as a function of $J_1$ and $J_2$) and a vertical plane (the first constraint) (Fig. \ref{classMinEPfig}). The minimum of the parabola gives the location of the stationary state. 

One can go further, and ask questions about the power supplied by the external voltage in the stationary state. One can use the energy balance equations, which in the stationary state looks like
\begin{eqnarray}
\frac{d\mathcal U}{dt}^*&=&\frac{dQ}{dt}^*-\frac{dW}{dt}^*=(U_1J_1+U_2J_2)^*-EJ_E^*=0, \label{Energybalance}
\end{eqnarray} 
with $\mathcal U$ the internal energy of the system (which is constant in the stationary state), $W$ the work input of the system and $J_E^*$ the stationary state current through the external voltage.
This means that according to the first law of thermodynamics the power supplied by the
external voltage, $dW/dt^*=J_E^*E$, is entirely dissipated into heat: $dW^*/dt=dQ^*/dt$. This 'all work is dissipated' law is only valid in the stationary state, and can be used as a second constraint to calculate $J_E^*$ or $\frac{dW}{dt}^*$. In the near-equilibrium MaxEP principle this second constraint will be of higher importance.

Some remarks: 
\begin{itemize}
\item In this example, there is one independent varying force $U_1/T_0$ and one fixed force $E/T_0$. The other dependent varying force is proportional to $U_2=E-U_1$. The
heat production can be written as $dQ/dt \equiv U_1(J_1-J_2)+EJ_2$. The flux $J_1-J_2$ conjugate to the varying force vanishes indeed in the
stationary state. 
\item This MinEP only works for equal temperatures. For unequal temperatures the currents (of order $E$) differ at order $\Delta T$, so the error is order $E\Delta T$. For the electrical network, minimizing the heat production (MinHP) instead of entropy
production, however, gives a better result, valid at higher order. 
\item Using the dynamics (\ref{dynJ}), it can also be shown that $d\sigma/dt\leq 0$. 
\item It should be stressed that this MinEP principle is superfluous once all balance and constitutional equations are known. The usefulness of this principle lies in the possible determination of the stationary state if not all the balance equations, e.g. the electrical charge balance equation, are known. 
\item Besides the above macroscopic description, there is also a microscopic versions of the MinEP principle. Proofs were given by e.g. Prigogine, and in the microscopic formulation for time-reversible even degrees of freedom in e.g. \cite{MaesNetocny, Klein}. However, as e.g. Landauer \cite{Landauer} pointed out, this MinEP principle is not valid in e.g. a linear circuit with a resistance in series with an inductance. It can be shown \cite{BruersMaesNetocny} that this is due to the fact that one has to deal in this case with currents instead of charge densities as basic variables, and these currents are time-reversible odd degrees of freedom.
\item As the Least Dissipation principle for the transient regime was related with fluctuation (Large Deviation) theory, also the MinEP principle for the stationary states can be related in an equal way, with the help of a Lagrangian and a probability measure on path space. This means that with the help of MinEP one can also describe the behavior of fluctuations around the stationary state \cite{BruersMaesNetocny}.
\end{itemize}

\noindent \textbf{To summarize}: The MinEP assumptions and prerequisites are: 
\begin{itemize}
\item a linear relationship between forces and currents (like Ohm's law), with known constant
(not dependent on the currents, the forces or the time) resistances and external forces,
\item heat production as the sum of products of forces and currents,
\item equal temperatures (but this is unnecessary in case of MinHP),
\item the network structure should be known,
\item Kirchoff's loop law as constraints (one for each independant loop),
\item the 'all work is dissipated' stationary state law,
\item the dynamics have only one unique stationary state,
\end{itemize}
and it is used to find this unique stationary state, which can also be derived from the complete energy and charge conservation laws. The latter conservation law is nothing but
Kirchoff's law of currents. So the unique stationary state is the state with a lower EP than the other states satisfying the constraints. The principle is not true for e.g. systems in the non-linear-response regime, and also for linear systems with time-reversible odd degrees of freedom.

\section{Near-equilibrium MaxEP}

The near-equilibrium MaxEP principle is also a variational principle to find the unique
stationary state near equilibrium (i.e. in the linear response regime). 
It was used recently for e.g. electrical networks in \cite{Zupanovic1, Zupanovic2}, but a much older version of this principle already appeared in e.g. \cite{Ziman}. Dewar \cite{Dewar2} has worked out a proof by using a microscopic path-informational entropy maximization principle (however, contrary to its claim,
the proof in \cite{Dewar2} only works for near-equilibrium systems). An idea similar to MaxEP can also be applied to find the stationary states for other dynamical systems, where an
EP-like function is constructed and maximized.
An example is a generalized Lotka-Volterra dynamics \cite{Jorgensen}. And as was noted in \cite{Christen}, one can derive the MinEP principle by a 'contraction' of a system with MaxEP.

\begin{figure}[!ht]
\centering
\includegraphics[scale=0.7]{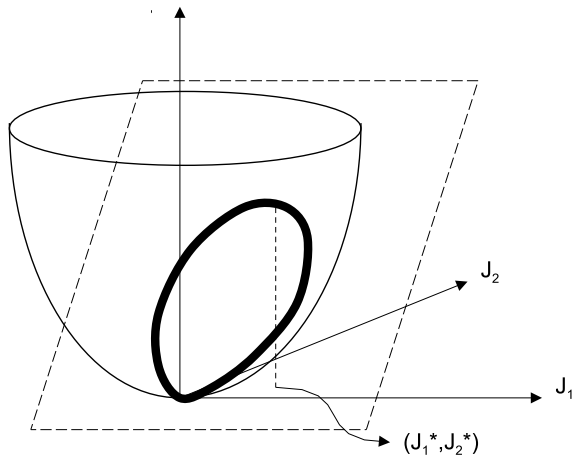}
\caption{Schematic plot of the MaxEP principle. The thick ellipse is the intersection of the quadratic EP potential with the tilted plane through the origin (dashed lines) as constraint. The maximum of this ellipse gives the stationary state currents. The other points on this ellipse are \emph{not} physical, transient states.}
\label{classMaxEPfig}
\end{figure}

To explain MaxEP in the electrical network example (based on \cite{Zupanovic1}), consider a circuit with two resistances $R_1$ and $R_2$, but
this time they are placed in parallel. The external fixed driving force is again given by the voltage $E$, the stationary state electrical current through this voltage is
$J_E^*$ and the currents through the resistances are $J_1$ and $J_2$. The expressions for the heat production, the power input, the dissipation function and the EP are the same as in
the MinEP example. Again, for MaxEP but not for MaxHP\footnote{So the principle in \cite{Zupanovic1} is better called MaxHP}, we have to make
the equal temperature restriction and also the linear relation between forces and fluxes (i.e. the resistances are constants). In contrast to
MinEP, we now assume for the stationary state Kirchoff's law of currents instead of the loop law, so this means $J_E^*=J_1^*+J_2^*$. To find the unique stationary state, the two kinds of constraints are this current law (charge conservation) and the 'all work is dissipated':
\begin{eqnarray}
J_E=J_1+J_2 \label{chargeconsconstr}
\end{eqnarray} and
\begin{eqnarray}
dW/dt=dQ/dt. \label{awidconstr}
\end{eqnarray}
Remark that these are also used outside the stationary state, even though they are not valid in the transients (even worse, the external current is not necesseraly well defined in the transients). With Lagrange multipliers, the following function
\begin{eqnarray}
\mathcal L\equiv \sigma +\lambda_1 (EJ_E-R_1J_1^2 - R_2J_2^2)+\lambda_2(J_E-J_1-J_2)
\end{eqnarray} 
is to be extremized. Its solution is $J_1^*=E/R_1$ and $J_2^*=E/R_2$, and this gives
Kirchoff's loop law. It can be visualized as follows (Fig. \ref{classMaxEPfig}). The EP is again a quadratic function of the two currents $J_1$ and $J_2$, which is a
paraboloid. The 'all work is dissipated' constraint $E(J_1+J_2)-R_1J_1^2 - R_2J_2^2$ gives an elliptic intersection with the EP paraboloid, with one extremum (minimum) for
$J_1=J_2=0$ and one maximum for the stationary state values.

Some remarks:
\begin{itemize}
\item There is no contradiction between MinEP and MaxEP. They use different constraints and assumptions.
\item An important difference between MinEP and MaxEP is that in MaxEP, we do not have $d\sigma/dt \geq 0$, and that all of the constraints are stationary state properties extended
out of stationary state, whereas in MinEP the Kirchoff loop law constraint is also valid in the transients. So the extension of the first constraint (apply it for the transients as well) in MaxEP is in a sense less physical than in the MinEP principle. We can therefore conclude that MinEP is a stronger principle than MaxEP, because the former also gives information about the time behavior. 
\item In the system with parallel currents, $E$ is fixed, whereas the total current $J_E^*$ through this external voltage is variable, but subject to two constraints (\ref{chargeconsconstr}) and (\ref{awidconstr}). Let us now take the opposite case: fixed current $J_E$ and variable but constrained voltage $E$. Solving this system gives MinEP, and is the complete dual description of the MinEP in the series network.
\item Applying MaxEP to the electrical circuit with two resistances in series or MinEP to
the parallel circuit becomes trivial: there is nothing left to vary because there are to much constraints.
\item As will be explained in \cite{BruersMaesNetocny}, this MaxEP principle. like the Least Dissipation and the MinEP, is also related with fluctuation theory.
\end{itemize}

\noindent \textbf{To summarize}: The near-equilibrium MaxEP assumptions are: 
\begin{itemize}
\item a linear relationship between forces and currents (like Ohm's law), with known constant
(not dependent on the currents, the forces or the time) resistances and external forces, i.e. $E$ and $R_i$ are fixed,
\item heat production as the sum of products of forces and currents.
\item equal temperatures (but this is unnecessary in case of MinHP).
\item the network structure should be known,
\item the 'all work is dissipated' law extended out of the stationary state as first constraint,
\item Kirchoff's law of currents extended out of the stationary state as second constraint (one for all nodes),
\item the dynamics have only one unique stationary state,
\end{itemize}
and it is used to find the unique stationary state, which in this case can also be derived from energy conservation. The latter is nothing but
Kirchoff's loop law. So the stationary state is the state with a higher EP than the other states satisfying the constraints. The principle is not true for non-linear-response regimes.

\section{Far-from-equilibrium non-variational MaxEP} \label{FFE NV MaxEP}

This relatively new principle is used in many different contexts. In order to avoid confusion, we first need to explain a bit what is meant with far-from-equilibrium. The far-from-equilibrium notion here denotes far-from-global-equilibrium with non-linear-response laws, but the framework of local thermodynamics is still assumed to be a good approximation. So it assumes a deterministic, non-linear dynamics of macroscopic variables without linear force-flux relationships. The non-linear dynamics should not be confused with the non-linear force-flux (i.e. the linear response) relationships: Close to global equilibrium (linear response regime) systems can have non-linear dynamics in their balance equations \cite{Demirel}. And also non-linear response systems can have linear dynamics. A related issue is the possible dependence of the linear response matrix $L_{ij}$ on the forces (like temperature gradients), but also on other dynamic quantities (like temperature fields,...). In this section, with the explicit example below, both dynamics and force-flux relations are non-linear, and the (non-)linear response matrix explicitly depends on variables. 

The principle works as follows: The dynamics are solved in order to find the (multiple)
stationary states. The question is: Which of these stationary states is physically selected, i.e. 'most real'? One can arrange the states in
increasing order from the least stable to the most stable. The notion of stability can be understood as e.g. linear stability around the stationary state. Another
possibility is to classify the states according to the relative size of their 'basin of attraction': the subset in phase space where all trajectories
starting in this set reach (asymptotically) the particular stationary state. The 'most likely' or 'most real' states are then defined to be the ones which are most stable, or which have the largest basin of attraction.

The MaxEP principle (hypothesis) states that the most real state is the one with the highest EP. An important remark that is not always
respected is that this principle is not a variational principle, because there is no variation with respect to continuous variables such as
fluxes. It is rather a selection principle of a discrete number of stationary states. In this sense, it is from a very different nature than the
far-from-equilibrium variational MaxEP (see below). It also does not claim that the EP always increases in time if the system evolves towards the 'most
real' stationary state.
Generalizations to periodic or chaotic attractor behavior is easily done by using time averaged EP and other notions of stability. 

The author is not
aware of microscopic or stochastic descriptions of this MaxEP principle. There is no general proof of this principle, and in fact, it can easily
proven to be wrong in a lot of cases, as is shown in e.g. \cite{Bruers2007}. Since non-variational MaxEP is not always valid, the search is for criteria for the situations where it is valid. What are the restrictions on the dynamics or the response matrix, in order that MaxEP is valid? Compare it with the near-equilibrium MinEP principle, where the criterium is the linearity of the response matrix.

This MaxEP principle is also related to the notion of dissipative systems with dissipative structures. If one drives the system out of equilibrium, at
certain critical points of the driving force, bifurcations to other stable states are possible. Then a patterned or ordered structure might
arise. A famous example is the Rayleigh-B\'enard system \cite{Rayleigh}. This consists of a viscous fluid subject to a gravitating field and a
temperature gradient: The bottom layer is heated whereas the upper is cooled. At a critical level of the temperature gradient, the
heat-conducting state is transformed to a heat-convecting state, with convection cells in a regular pattern, called the dissipative structure. 
The claim is that these ordered dissipative structures (if they exist) always have a higher EP than the so called 'thermodynamic branch' states without the structures, i.e. the states which do
not show a pattern, like the conduction state.

The applications in the literature are frequent: they are applied to turbulent and (oceanic) fluid systems \cite{Shimokawa, Ozawa}, to B\'enard convection or to autotrophic
ecosystems \cite{Schneider}, to the morphology of crystal growth \cite{Martiouchev}, to chemical reaction systems like the Brusselator or to
growing random surface patterns \cite{Sawada, Sawada2} or to heterotrophic ecosystems \cite{Bruers}. This principle can also be found in \cite{Woo}, but there it is somewhat confused with the least dissipation principle.  Most of these authors restricted
their description to the dissipative structures after the first bifurcation. 

The situation is more complicated as one drives the system even further out of equilibrium, where new bifurcation points might appear and
transitions to other stable (or periodic, chaotic,...) states occur. It is not guaranteed that these new dissipative structures always have an EP higher than the former
structures, as is explicitly shown in the example of a chemical reaction system in \cite{Bruers2007}. This result that the most stable state is not necessarily the one with highest EP after a second bifurcation, is also seen in the convective fluid system, as in Nicolis \cite{Nicolis} or Clever \cite{Clever}. In the latter work, the heat transport near the convective roll instability was studied, and it was shown that the heat transport (which is proportional to the EP) of the stable wavy rolls was lower than the transport of the unstable straight rolls.

\textbf{To summarize}: The far-from-equilibrium non-variational MaxEP assumptions and unknowns are:
\begin{itemize}
\item the expression for the EP,
\item a number of steady states, resulting from a non-linear dynamics, are known to exist,
\end{itemize}
and it is a principle to select the 'most real' state out of this number of stationary
states. The principle is not always valid if 'most real' means 'most stable' or 'largest basin of attraction'. No correct criteria for systems where this MaxEP does apply are proposed.

\section{Far-from-equilibrium variational MaxEP}

The far-from-equilibrium variational MaxEP principle is perhaps the most interesting one. There are no satisfying theoretical proofs of it, and it is
clearly not always correct. But there is an increasing amount of (apparently) experimental verification for specific interesting systems. First, with the help of an electrical circuit example, an attempt is made for a correct statement of this principle. Afterwards a rough overview of the literature (without trying to be complete) is presented. 

The correct statement of this principle is not easy, because it largely depends on what is known
and presupposed, and it also depends on the e.g. the relevant time-scales. In general, there are a number of
resistances in the circuit, but now not all the values of these resistances are known. This makes the principle totally different from the former ones,
because previously the behavior of all resistances was sufficiently known. More concretely: Consider a circuit with two resistances (plus parallel capacitances) placed in series. The potentials along the circuit are $V_H$ (constant), $V_1(t)$ (variable) between the resistances, and $V_L$ (constant). The fixed external force is given by  $E=V_H-V_L$, and the external current $J$ is not fixed (systems with variable external forces and fixed externally supplied currents are also possible for this MaxEP). Suppose the first resistance $R_1$ is known (and constant), as well as its functional relation between forces and currents. As an example, we take the linear approximation,
\begin{eqnarray}
J_1(t)=(V_H-V_1(t))/R_1, \label{J1=(VH-V1)/R1}
\end{eqnarray}
The linear approximation of this first resistance does not necessarilly implie a close to global equilibrium approximation, because the underlying dynamics as well as the force-flux relation at the second resistance might be highly non-linear. We have called it far-from-equilibrium because this MaxEP principle appears to be mostly valid (if it is experimentally valid after all) in far-from-equilibrium (non-linear) systems, like the climate system.
The dynamics or value of the second resistance, nor its functional relationship between the forces and fluxes at this resistance are known. A priori, the resistance might depend on say the current $J_2$ flowing through the resistance, or it can even depend on some new or unknown variables. To proceed, one has to make exact statements of the principle, and there are different formulations of the problem: a deterministic and a stochastic one.

A first, deterministic formulation is that in certain systems there is a unique state showing a deterministic negative feedback mechanism due to a trade-off mechanism at the second resistance, and leading to a maximization of the EP at this resistance. This trade-off goes as follows: On the one hand, this resistance wants to increase its
current through it, but as the current becomes to large, the voltage drop over this resistance becomes to low. If the resistance
also prefers a large voltage drop (a large force), there is a trade-off between large force or large flux, and it might evolve or 'self-organize' itself to optimize this
trade-off, which means the largest EP at this resistance is taken. 

The state of MaxEP at the second resistance might also arise in a probabilistic way. Just like the maximum entropy principle in equilibrium thermodynamics is based on a counting argument in statistical mechanics using a 'thermodynamic limit', i.e. using the law of large numbers for systems with a lot elementary states or degrees of freedom, the variational MaxEP might perhaps also be derived in a kind of thermodynamic limit. The author is not aware of a clear derivation, and attempts like the ones made by Dewar \cite{Dewar1, Dewar2} are not yet fully convincing (its shortcomings will be presented in future work). In the probabilistic setting, there are at least two possibilities.\\
-The system has enough degrees of freedom in order to have a very high amount of stationary states. Perhaps most of these states are clustered around the values for maximum EP at the second resistance. This amount to taking a probability measure on the stationary states.\\
-The system has enough microscopic degrees of freedom that the macroscopic stationary state corresponding with MaxEP at the second resistance has the most microscopic trajectories, i.e. this amounts to taking a probability measure in path space.

Of course, different descriptions might be equal, or there might exist other
mechanisms. These formulations are still quite vague and unexplicit and should be interpreted as guiding lines to find concrete non-trivial
models where MaxEP is realized. The author believes that the search for such models is a major challenge.

There are a few points that make the correct formulation a bit tricky. First, the above description was in terms of stationary states. But perhaps the principle should also be formulated for periodic or chaotic states, because most real systems (like the climate) are apparently not in a well defined stationary state. One could use time-averages of the EP, but this off course becomes tricky if multiple timescales are
present, as is often the case in far-from-equilibrium systems. In most 'self-organizing' systems, there is this large difference in timescales. For example: In ecosystems, one has highly different timescales, corresponding with e.g. behavior of individuals for short timescales, population dynamics for longer ones and evolution and natural selection for even longer timescales. In the electrical circuit description, one can also state that there is a vast difference in the short timescale of microscopic electron motion, and a slow evolution of 'self-organizing' resistances. So how do we know at what timescale the principle works? Perhaps it is possible to find correct criteria, and that one finds that the parameters in the short timescale dynamics (like the unknown resistances) are not really constants, but are slowly varying variables, evolving towards the correct values for MaxEP.

A second unclarified point appears when there are multiple unknown resistances. Some problems might arise about the compatibility of
maximizing all the EP's at these resistances simultaneously. It is important to note that there is some
arbitrariness about which stationary state EP's are to be maximized: the total EP or some combinations of specific EP's at specific resistances.

For the moment, we will not go into these unclarified points, and we are only interested in the stationary state (or time averaged) EP, and only in the EP at the unknown resistances, not the total resistance.
Returning back to the electrical circuit example: The EP in the stationary state from the second resistance is given by
\begin{eqnarray}
\sigma_2^*=J_2^* X_2^*=J_2^*(V_1^*-V_L)/T_2,
\end{eqnarray}
whereby we know that in the stationary state $J_2^*=J_1^*$. With (\ref{J1=(VH-V1)/R1}) and this stationary state condition we get
\begin{eqnarray}
U_2^*=V_1^*-V_L=V_H-V_L-J_2^*R_1=E-J_2^*R_1.
\end{eqnarray}
Then we have $T_2\sigma_2^*=J_2^*(E-J_2^*R_1)$. Take now the variation of this stationary state EP with respect to $J_2^*$ and it gives as solution $J_{max}^*=J_{1,max}^*=J_{2,max}^*=E/(2R_1)$. It can easily be shown to be a maximum, so therefore it is called MaxEP. This gives for the second resistance $R_{2,max}^*(J_{i,max}^*,V_{j,max}^*,...)\equiv U_{2,max}^*/J_{2.max}^*=R_1$. This derivation is completely analogous with the maximum power theorem in electrical systems and control thermodynamics \cite{Jackson, Salamon}, which states that to obtain maximum power from a source with a fixed internal resistance (the first resistance), the resistance of the load (the second resistance) must be made the same as that of the source. In \cite{Bruers2007} this principle was called the partial steady state MaxEP, beacuse the variational quantity is the partial (not the total) EP in steady state conditions.

Having described the principle, let us now briefly overview the literature. The starting point of this MaxEP approach was due to work by Paltridge \cite{Paltridge1, Paltridge2}, to understand the convective heat flows in the earth atmosphere. The idea is that the heat transport coefficient (like the
resistance) of the atmosphere, the heat flow (the electric current through this resistance), and the driving force (temperature gradient, the
potential over the resistance) on earth are not fixed, and that they will settle themselves in a state of MaxEP. This approach has been made more precise and extended to other planets \cite{Wyant, Ohmura, Grassl, Lorenz1, Lorenz2, Kleidon}. Also mantle-core convection has been studied in this respect
\cite{Vanyo}. These are experimental verifications (we refer to the cited work for the experimental methods used) and the author is not aware of
any satisfying theoretical underpinning. The difference of these climate systems with the above described electrical network system is that in the electrical network we had a fixed external voltage E, and a variable flux $J$, whereas in the climate systems there is a fixed external flux $J$ (the solar radiation constant), but variable external driving forces (because they depend on the variable earth temperatures).

Besides the possibilities in atmospheric and fluid systems, also in ecological systems this MaxEP principle might appear. Starting with the above mentioned maximum power theorem, Lotka, Odum and others \cite{Odum, Hall} extended this maximum power principle to ecosystems. However, whether this extension to ecosystems is really the variational MaxEP principle, and the status of experimental verifications is unclear to the author. There are rough experimental indications that this MaxEP might work for specific ecosystems like sediment ecosystems at the bottom of the sea (this will be presented in future work \cite{Bruers}).

Another approach is to accept MaxEP and use it as a modeling technique. This is done e.g. in \cite{Zupanovic3} for photosynthetic chemical
reactions, and this is also compatible with experimental data. Furthermore it is applied to model self-organizing climates and Gaia-models
\cite{Toniazzo}, but this is criticized by \cite{Ackland} who showed that the daisyworld Gaia-model did not evolve to the MaxEP state. On the other hand, Kleidon \cite{Kleidon2} argued that there is a correspondence between MaxEP and Gaia: The biosphere evolves towards the MaxEP state and since this state is in homeostasis due to negative feedbacks, the biosphere creates a homeostatic state 'by and for itself', which is the vague formulation of the Gaia hypothesis.

The
biggest challenge in theoretical MaxEP research is perhaps to find a non-trivial model where the system approaches or settles itself in the MaxEP state. This should give more
information why MaxEP works for a lot of systems like convective fluid systems or perhaps also ecosystems. Some very rough ideas, which are still far from satisfying, can be found in \cite{Kleidon} or \cite{Paltridge3}. Questions like, how much degrees of freedom are needed, what are the criteria for the timescales, are there dynamical or kinetical constraints necessary in order that MaxEP is valid, is it a trade-off or a statistical consequence... are not answered yet. Or perhaps the MaxEP experimental verifications are wrong and only a coincidence, and further research is useless?

\textbf{To summarize}: The far-from-equilibrium variational MaxEP assumptions and unknowns are: 
\begin{itemize}
\item The system is in the stationary state, and the stationary state EP from the known and unknown resistances is given by the product of currents and forces, 
\item from the stationary state condition, Kirchoff's law of currents is known,
\item some resistances $R_k$ are known, as well as their functional relation between the forces and the currents (like Ohm's law),
\item the external fixed voltagedifference $E$ is known,
\item no equality of temperatures is required,
\item for the unknown resistances $R_u$, a relation between the forces and the unknown currents is not necessarily known,
\end{itemize}
and the stationary state EP's (given as a function of the known force $E$, the known parameters $R_k$ and the unknown stationary state currents $J_u^*$) at the unknown resistances are maximized with respect to the unknown currents. This gives $J_{u,max}^*$ as a function of $E$ and $R_k$. So these special values of $J_{u,max}^*$ result in a higher EP compared to other $J_u^*$ values. This principle is evidently not valid for every system, but some systems appear to have experimental verifications. The interesting search is to find theoretical criteria or mechanisms in systems where it does apply. 

\section{Optimization MinEP}

This MinEP principle is very different in nature from the other principles, but is added for completeness sake. It is used to study efficiency in engineering systems, and studied in so called control thermodynamics \cite{Salamon}. The above principles all involved 'uncontrolled' processes whereby the system itself will do the minimization or maximization. In the optimization principles, it is the experimenter who is able to control and will have to choose the correct voltages or resistances in order to extremize a quantity. 

Let us apply this technique to a simple electrical network system. Therefore, consider again a network with external environmental voltage $E$, and the system now consists of two resistances $R_1$ and $R_2$ placed in parallel. The idea is that one now wants to control the voltage drops over the resistances. Therefore, we extend the system by introducing two other external voltages $E_1$ and $E_2$, its values controlled by the engineer, placed in series with the two resistances respectively. The total current (in the stationary state) through the external voltage $E$ is e.g.: $J_{tot}=(E+E_1)/R_1+(E+E_2)/R_2$. This is the linear response (Ohm's law) approximation for convenience, but naturally the formalism works also for non linear behavior and far-from-equilibrium systems.

Now the engineer wants to take the best voltages $E_1$ and $E_2$ in order to attain maximum efficiency, keeping the value of the total current constant. There are different notions of efficiency: maximize power, minimize loss of work (minimum exergy degradation),... Let us take minimum EP as the efficiency measure and the total current as constraint. Therefore we again use the Lagrange multiplier method and take the variation of
\begin{eqnarray}
\sigma+\lambda J_{tot}&=&\frac{(E+E_1)^2}{T R_1}+\frac{(E+E_2)^2}{T R_2}+\lambda \left(\frac{E+E_1}{R_1}+\frac{E+E_2}{R_2}\right),
\end{eqnarray}
with respect to $E_1$ and $E_2$. The solution gives $E_1=E_2=J R_1 R_2/(R_1+R_2)-E$. So the controllable forces should be equal, and therefore this principle is called the equipartition of forces \cite{Sauar}.

One can imagine that also other systems, like biological systems, experience the same challenge as an engineer of maximizing its efficiency by regulating system forces instead of internal resistances. So perhaps this principle is also usefull in nature?

\textbf{To summarize}: the optimization MinEP assumptions and unknowns are:
\begin{itemize}
\item EP, constitutive equations and internal resistances are known,
\item the total current is set fixed,
\item the system voltages are unknown,
\end{itemize}
and the EP is minimized with respect to the unknown system voltages in order to find the best values of these voltages. This principle is evidently always true.

\section{Other proposals and further discussion}

\begin{figure}[!ht]
\centering
\includegraphics[scale=0.7]{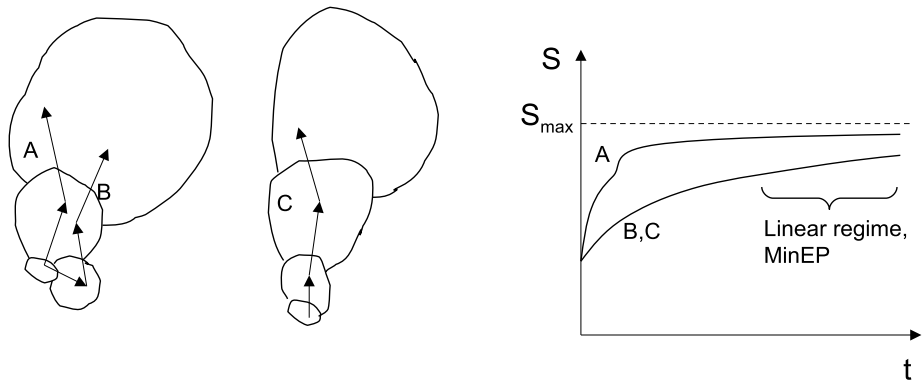}
\caption{In phase space there can be different paths. Path A starts far-from-equilibrium, and corresponds with the MaxEP. However, it is not excluded that path B is followed, or perhaps the structure of the compartments in phase space is different, and path C is followed. Paths B and C will have a lower EP. The right figure shows the time dependence of the entropy $S$. In the stationary state of the linear regime, there is always MinEP consistent with the constraints.}
\label{otherMEPfig}
\end{figure}

There are some other proposals, but however, these are not always clarified or correct. We will briefly list some extra principles, statements or formulations that one might encounter in the literature. A big puzzle remains whether these different formulations are refering to different principles or not. And are they related with the abovementioned principles? Let me stress that the discussions in this section are still (highly) speculative.
\begin{itemize}
\item It is very often claimed \cite{Martyuchev, Sawada, Swenson, Woo} that a non-equilibrium system (not necessarily in the stationary state) relaxes to equilibrium as fast as possible. The idea behind this is that a non-equilibrium system is in a small region in phase space with low entropy, and has the highest possibility to evolve in the next time step to the largest region, with highest entropy. However, this argument is fallacious, because the dynamics might not allow such evolutions, or the low entropy and high entropy phase space regions can be highly separated (see Fig. \ref{otherMEPfig}). The 'as fast as possible' statement involves constraints due to the microscopic dynamics, and hence the usefullness of this MaxEP statement is dubious (especially if the dynamics is deterministic, because given the initial conditions, there is only one path possible). It also depends on our knowledge of the system, whether we know the phase space point, or whether we only have a propability distribution. Macroscopically, these remarks translate themselves to the fact that the relaxation time of the syetem is non-zero.
\item Kleidon discussed in e.g. \cite{Kleidon2} another principle applied to planetary climate and ecosystems. There is a constant inflow of solar radiation at the earth, but part of this low entropy radiation is reflected. However, it is claimed that biological systems evolve such as to capture more efficiently this incoming radiation, turning it into high entropy heat radiation. So the ecosystem lowers the planetary albedo, and the highest entropy production is reached when the albedo is zero. The earth with zero albedo becomes a black body for short wavelengths and as is shown in \cite{Santillan}, the black-body radiation is indeed in the MaxEP regime. Due to the fact that the earth albedo is not (yet) zero, this MaxEP principle is not (yet) true. Although this principle is also proposed for climate systems, it should not be confused with the variational MaxEP principle.
\item There is a Min-MaxEP principle due to Aoki. This suggests that in some ecosystems or biological systems the EP in function of time first increases towards a maximum, and later on decreases again (as the climax is reached). There are no theoretical proofs of this principle, and the experimental verification is still dubious.
\item The far-from-equilibrium near-steady-state MaxEP principle. The idea is that for systems far enough from equilibrium, near the steady state(s), the (total or a partial) EP increases: $d\sigma/dt \geq 0$. So the EP is like a local Lyapunov function. It is the same kind of principle as the near-equilibrium MinEP, but with the opposite time behavior of the EP. However, no satisfying criterium for the required minimal distance from equilibrium is given. Nevertheless, recently \cite{Christen} there was a nice argument in favor of this principle with the example of the Steenbeck principle for electric arcs. It is somewhat related with the far-from-equilibrium variational MaxEP: Suppose that the second resistance $R_2$ in the series network example depends on some variables $z$. It is possible that the system will settle in the state with values $z^*$, such that the total EP is maximal (compared to the EP for other values of $z$). The steady state of a gas discharge system also minimizes the voltage over the second resistance, something which has experimental verification (see \cite{Christen} for more references). Also other authors might implicitly discuss about this principle (perhaps in e.g. \cite{Johnson, Schneider, Sawada} or others), or their work can be interpreted in this way. This principle is not true in the explicit example discussed in the far-from-equilibrium non-variational MaxEP section, nor in Aoki's Min-MaxEP or others. 
\item In \cite{Jorgensen} are some suggestions for MinEP related principles for ecological and biological systems. The author is very critical about these principles, and comments on them will be given in future work.
\item With statistical mechanics arguments, one can also formulate MaxEP principles. \v Zupanovi\'c \cite{Zupanovic4} proposed the idea that given a probability measure in path space, the path with highest microscopic (path-dependant) EP has the highest probability. This is equivalent with the statement that in the canonical Gibbs ensemble, the state with the lowest energy has the highest probability. However, it is possible that there are much more paths with low EP, leading to the difference between the path-averaged, macroscopic EP and the microscopic path-dependant EP. Furthermore, this 'statistical mechanics' MaxEP principle is based on the supposed property that the path action is proportional to the EP. But this is not always the case. E.g. \cite{Maes} and \cite{BruersMaesNetocny} discuss that the time-antisymmetric part of the action equals the microscopic EP, but in most cases there is also a time-symmetric part which might depend on the microscopic path.
\end{itemize}

\section{Conclusions and directions for future research}

In this article, some comments on different macroscopic entropy production principles were given and a classification was made between variational and
non-variational principles, between MinEP and MaxEP, between near-equilibrium and far-from-equilibrium, between principles to find the
stationary states and principles to find the phenomenological laws (equations of motion) in the transient states, between experimentally tested or theoretically proven
principles, between different kind of constraints and between optimization and uncontrolled system principles. With this setting, a (far from complete) overview was made of the relevant literature, in order to disentangle some
approaches. 

There is a lot of confusion and misinterpretation in the literature. The author hopes that this classification and these comments will help further the
research, without declining into vague statements and concepts. The biggest challenge lies perhaps in the search for a theoretical non-trivial
example where far-from-equilibrium variational MaxEP applies. This might help to understand more deeply the functioning of atmospherical,
hydrological, biological, ecological or other self-organizing systems. Other interesting research directions are:\\
-the search for microscopic and stochastic foundations of some EP extremization principles, \\
-the search in the mesoscopic area \cite{Reguera} and the far-from-equilibrium and non-local regime (which is of high importance for a better understanding of biological systems), \\
-the attempt to formulate criteria of validity for proposed principles, \\
-or trying to see whether the proposals in the last section have relations with others.

\section*{Acknowledgments}

The author wishes to thank F. Meysman, C. Maes, R. Dewar, P. \v Zupanovi\'c, K. Neto\v cn\' y, T. Jacobs, W. De Roeck and T. Christen for helpful discussions and comments.


\end{document}